\documentclass[pdflatex,sn-basic]{sn-jnl}

\usepackage{adjustbox}
\usepackage{placeins}
\usepackage{subcaption}
\usepackage{amsfonts}
\usepackage{physics}

\makeatletter 
\def\@acknow{}%
\long\def\EarlyAcknow#1 \par{%
\def\@acknow{\abstractfont\abstracthead*{Acknowledgments}
#1\par}}%

\def\printabstract{\ifx\@acknow\empty\else\@acknow\fi\par%
    \ifx\@abstract\empty\else\@abstract\fi\par}
\makeatother

\jyear{2021}%

\theoremstyle{thmstyleone}%
\theoremstyle{thmstyletwo}%

\theoremstyle{thmstylethree}%

\begin{document}

\title[Classical Ensembles of Single-Qubit Classifiers]{Classical Ensembles of Single-Qubit Quantum Variational Circuits for Classification}


\author*[1,2]{\fnm{Shane} \sur{McFarthing}}\email{shane.mcfarthing@gmail.com}

\author[1,3]{\fnm{Anban} \sur{Pillay}}

\author[1,2]{\fnm{Ilya} \sur{Sinayskiy}}

\author[4,2]{\fnm{Francesco} \sur{Petruccione}}

\affil*[1]{\orgname{University of KwaZulu-Natal}}

\affil[2]{\orgname{National Institute for Theoretical and Computational Science (NITheCS)}}

\affil[3]{\orgname{Centre for Artificial Intelligence Research (CAIR)}}

\affil[4]{\orgname{Stellenbosch University}}


\EarlyAcknow{We would like to thank the National Institute for Theoretical and Computational Sciences (NITheCS) for funding this research, as well as the Centre for Artificial Intelligence Research (CAIR) for providing equipment for testing.}

\abstract{The quantum asymptotically universal multi-feature (QAUM) encoding architecture was
recently introduced and showed improved expressivity and performance in
classifying  pulsar stars.   The circuit  uses  generalized trainable layers of
parameterized single-qubit rotation gates and single-qubit feature encoding
gates. Although the improvement in
classification accuracy is promising, the single-qubit nature of this
architecture, combined with the circuit depth required for accuracy, limits its
applications on NISQ devices due to their low coherence times.  This work
reports on the design, implementation, and evaluation  of  ensembles of single-qubit QAUM classifiers using classical bagging and
boosting techniques.   We demonstrate an improvement in validation accuracy for
pulsar star classification. We find that this improvement is not problem
specific as we observe consistent improvements for the MNIST Digits and
Wisconsin Cancer datasets.  We also observe that the boosting ensemble achieves
an acceptable level of accuracy with only a small amount of training, while the
bagging ensemble achieves higher overall accuracy with ample training time.
This shows that classical ensembles of single-qubit circuits present a new approach for certain classification problems.}




\maketitle

\section{Introduction}\label{section:1}

The potential to efficiently run classically-intractable algorithms on quantum devices has driven the development of both quantum algorithms and quantum computers themselves, with the goal of obtaining a fully-fledged quantum device with error-mitigation \citep{nielsen}. Quantum computers utilize the principles of quantum mechanics to operate on data and are built on the foundational concept of the qubit, a quantum bit which unlike the classical bit, can also exist in a superposition of the states 0 and 1 \citep{nielsen}. Additionally, quantum entanglement allows the qubits in a quantum computer to be correlated in a manner not possible in classical computing, resulting in a significant increase in computational power due to the possibility of parallel operations on multiple qubits \citep{nielsen}.

The realization of full-scale quantum computers on which it is possible to perform error-correction to deal with noise is a significant engineering challenge \citep{petruccione}. Implementing robust error-correction schemes requires quantum devices possessing many orders of magnitude more qubits than the quantum computers available today \citep{petruccione}. As a result, noisy intermediate-scale quantum (NISQ) devices present a viable platform for the development of new quantum algorithms and techniques for mitigating errors due to noise \citep{nielsen}. These NISQ devices have much lower numbers of qubits and cannot run error-correction routines to combat noise, however they are still capable of performing tasks that are difficult on classical computers, such as the simulation of quantum systems \citep{petruccione}.

 Many near-term algorithms have been developed to run on these NISQ devices as a way to leverage the hardware available today. One particular class of algorithms that can efficiently utilize NISQ devices are hybrid quantum-classical algorithms \citep{petruccione}. These algorithms prepare and run parameterized quantum circuits (PQC's) on quantum computers and use the output to optimize the parameters of the PQC on classical computers \citep{nielsen}. Some examples are the quantum approximate optimization algorithm (QAOA) which can be applied to combinatorial optimization problems, and the variational quantum eigensolver (VQE) algorithm for computing the ground state energy of a given Hamiltonian, with applications in nuclear physics and nuclear structure problems among others \citep{nielsen}. 

The field of quantum machine learning (QML) resulted from the application of quantum
computers to solve  machine learning problems.   QML involves the construction and training of mathematical models
capable of learning patterns and relationships in data and then applying these
to new data \cite{petruccione}. Many machine learning algorithms obtain the best solution to a problem using
optimization, and quantum machine learning makes use of the unique properties of
quantum systems to speed-up this optimization procedure \cite{petruccione}.
Despite the relatively recent emergence of this field, numerous quantum machine
learning algorithms have already been devised, such as quantum neural networks
(QNN) and quantum support vector machines (QSVM), and have shown promise in
various classical machine learning problems \cite{petruccione}. 

These quantum machine learning algorithms are mostly of a hybrid quantum-classical nature and make extensive use of PQC's \citep{petruccione}. The choice of circuit structure used in these algorithms is very important as it determines both the expressivity of the model \citep{Schuld} and its trainability \citep{mcclean}. As a result, the creation of PQC's with novel architectures capable of achieving high levels of generalization is an ongoing area of development \citep{schuld2015}.

One such novel architecture is the quantum asymptotically universal multi-feature (QAUM) encoding single-qubit circuit explored in \cite{mo}. This circuit design was inspired by \cite{Schuld}, which showed that a quantum model can be expressed as a partial Fourier series in the data, and that as more Fourier coefficients are made accessible through repetition of data encoding gates, the expressivity of the quantum model increases. The QAUM circuit makes use of generalized trainable layers of parameterized single-qubit rotation gates and single-qubit feature encoding gates and was used for the classification of pulsar stars \citep{mo}. When compared with the QAOA, the model built with the QAUM single-qubit circuit demonstrated the ability to obtain higher accuracy in its classification of the HTRU 2 Pulsar dataset  \citep{mo, htru}. Although this improvement in classification accuracy is promising, the single-qubit nature of this architecture and the circuit depth required for accuracy limit its applications on NISQ devices due to their low coherence times \citep{mo}.

In this work we explore the application of classical ensemble techniques with the QAUM classifier as means to scale up the QAUM architecture. We do this by utilizing multiple single-qubit QAUM circuits combined in an ensemble, rather than expanding the design of the QAUM circuit to multiple qubits. Though the concept using quantum ensembles is not new, with \cite{Schuld2018, macaluso} detailing techniques for creating ensembles of quantum states, the creation of classical ensembles of variational quantum classifiers is relatively unexplored. Ensembles of variational quantum classifiers that use classical voting strategies are explored in \cite{qin} and were shown to produce valuable improvements in classification accuracy over comparable quantum classifiers on IBM's NISQ devices. Classical bagging and boosting ensembles of quantum variational classifiers have also been explored in \cite{li2023ensemble}, which showed that higher accuracies and increased robustness are attainable using these ensemble techniques. Compared to wholly classical ensembles, quantum ensembles of quantum classifiers have been shown to have favourable scaling with regards to both ensemble size and training time. However, the number of qubits required to implement these quantum ensembles is currently out of reach of current-era quantum devices \citep{Schuld2018, macaluso}. 

The ensembles explored in this work utilize single-qubit circuits which can be executed on NISQ devices. Unlike \cite{qin} which uses plurality voting that is susceptible to classification errors on outlying datapoints, and \cite{li2023ensemble} which uses majority voting, we utilize a metalearner to obtain the optimal weighted voting strategy. Although both works \citep{qin, li2023ensemble} also investigate the use of classical ensemble techniques, they utilize generalized multi-qubit circuits, whereas we have specifically chosen the QAUM single-qubit circuit from \cite{mo}. This selection was made as our goal is to investigate the scaling of single-qubit architectures using ensemble techniques, rather than the more general use of ensemble techniques in \cite{qin, li2023ensemble}. 

The single-qubit nature of the classifiers in our work make the use of distributed quantum computing resources a possibility, as well as lower the requirements needed to run the ensembles on NISQ hardware. The ensembles use the QAUM classifier as their base learners and we test their classification performance on the HTRU 2 Pulsar dataset \citep{htru}. We find that the boosting and bagging ensembles outperform the single QAUM classifier from \cite{mo} when the amount of training permitted is both severely restricted and virtually unrestricted, respectively. We performed further testing on the MNIST Digits \citep{mnist} and Wisconsin Breast Cancer \citep{uci} datasets and verified that these improvements in performance were not limited to the pulsar classification problem.

The rest of this paper is organized as follows: in Section \ref{section:2} we briefly introduce the formalism of classification problems as well as the foundational work we build upon, in Section \ref{section:3} we detail the methodology used for the construction and testing of the ensembles, in Section \ref{section:4} we present the results obtained from the testing of the ensembles, and in Section \ref{section:conclusion} we give our conclusions and notes for areas of future work. Additionally, the graphs containing the results of each run of the cross-validation may be found in Appendices \ref{secA1} and \ref{secA2}.

\section{Background Information}\label{section:2}

In machine learning, an \textit{n}-dimensional binary classification problem can be described as a search for a model \textit{f} with \textit{i} parameters which maps \textit{n}-dimensional data to one of two classes, as shown below:

\begin{equation}
    f_{\boldsymbol{\theta}}(x) = y,
\end{equation}

where $\boldsymbol{\theta} = [\theta_1, \theta_2,\ldots,\theta_i]$ are the parameters of the model, $x \in \mathbb{R}^n$ is the input data, and $y \in \{0,1\}$ is the class label assigned by the model, with 0 and 1 being the binary class options. 

By encoding the data and parameters of a classification problem into quantum states and using quantum gates to operate on these states, we are able to create quantum classification models \citep{petruccione}. The classification of the input can be obtained by measuring the output of these PQC's, often referred to as quantum neural networks.

\subsection{Quantum Asymptotically Universal Multi-Feature (QAUM) architecture}

The QAUM single-qubit model comprises two circuit blocks: trainable layers made of parameterized single-qubit gates, and the multi-feature encoding gates \citep{mo}. For the encoding of the features of the dataset, each feature is scaled to lie in the range $(0, \pi)$ and used as a parameter for a \textit{Z}-rotation. Given an \textit{n}-dimensional feature vector, a single layer of the QAUM circuit contains \textit{n} encoding gates and \textit{$n+1$} trainable layers which are situated in an alternating fashion.

The expressivity of this architecture, by which we mean the classes of functions that it can learn, is a result of its capability to express the first degree multi-dimensional Fourier series of the dataset used \citep{mo}. All of the equations and derivations below have been taken from \cite{mo} and show that a truncated Fourier series of higher frequency can be obtained by increasing the number of repetitions in the circuit:

A circuit with a single encoding can be expressed in tensor notation as

\begin{equation}
\ket{\psi} = W^{(2)}S(x)W^{(1)}\ket{0} \to W^{(2)}_{ki}e^{i\mathcal{G}_{ij}x}W^{(1)}_{j1},
\label{equation:2}
\end{equation}

where $S(x)$ is the encoding used and $W$ are the weights of the trainable layers. In this case, we specify $S(x) = e^{i\mathcal{G}_{ij}x}$, with $\mathcal{G}$ being the Pauli-\textit{Z} matrix. Additionally, $\ket{0} \to [1,0]^T$ is absorbed into the second index of $W^{1}$ and the Einstein summation convention is used.

Generalizing to the case with $N$ features and $L$ encoding repetitions, and using the eigenvalues of $\mathcal{G}$, $\lambda \in \{-1,1\}$, we can rewrite Eqn \ref{equation:2} as

\begin{equation}
\ket{\psi}_k \to W^{(L^1)}_{ki_{L^1}}e^{i{\lambda_{i}}_{L^1}x_1}W^{(L^2)}_{i_{L^1}i_{L^2}} \ldots W^{(L^N)}_{i_{L^{N-1}}i_{L^N}}e^{i{\lambda_{L^N}x_N}}W^{((L-1)^N)}_{i_{L^N}i_{(L-1)^1}} \ldots W^{(0)}_{i_{1^N}1}.
\label{equation:3}
\end{equation}

The expectation value of a measurement operator, $\hat{M}$, in the $\ket{\psi}$ basis is obtained through a measurement of this qubit in the basis defined by $\hat{M}$, and is

\begin{equation}
\langle\psi\vert\Hat{M}\vert\psi\rangle = W^{\dag(0)}_{1j_{1^N}} \ldots W^{\dag(L^1)}_{j_{L^1}k'}M_{k'k}W^{(L^1)}_{ki_{L^1}} \ldots W^{(0)}_{i_{1^N}1}e^{i\sum_{l=1}^{N}x_l\left(\sum_{m=1}^{L}\lambda_{i_{m^l}}-\sum_{p=1}^{L}\lambda_{j_{p^l}}\right)}.
\label{equation:4}
\end{equation}

We can rewrite the exponent in Eqn \ref{equation:4} as

\begin{equation}
\beta = \sum_{l=1}^{N}x_l\left(\sum_{m=1}^{L}\lambda_{i_{m^l}}-\sum_{p=1}^{L}\lambda_{j_{p^l}}\right) = [1,1,\ldots,1]\left([\lambda_{i_{m^l}}]-[\lambda_{j_{p^l}}]\right)[x_1,x_2,\ldots,x_N]^T.
\label{equation:5}
\end{equation}

Through further simplification we obtain

\begin{equation}
\beta = 2\gamma[x_1,x_2,\ldots,x_n]^T,
\label{equation:6}
\end{equation}

where $\gamma = [\gamma_1,\gamma_2,\ldots,\gamma_N]$, with $\gamma_q \in \{-L,-(L-1),\ldots,-1,0,1,\ldots,L-1,L\}$.

This shows that through an increase in the number of encoding repetitions in the QAUM circuit, it is possible to approximate the classification problem with a truncated Fourier series of higher frequency \citep{mo}.

\subsection{Classical Ensemble Techniques}

Ensemble methods were developed as a way to combine the predictions of multiple machine learning models in a way that improves the overall performance of the system. It works by utilizing the diversity of the models in the ensemble to create a final prediction that is more robust and accurate than any one model in the ensemble \citep{dietterich}. The two types of ensemble methods that are explored in this work are bagging and boosting \citep{dietterich}.

\begin{figure}[t!]
\begin{adjustbox}{center}
    \centering
    \includegraphics[width=\columnwidth]{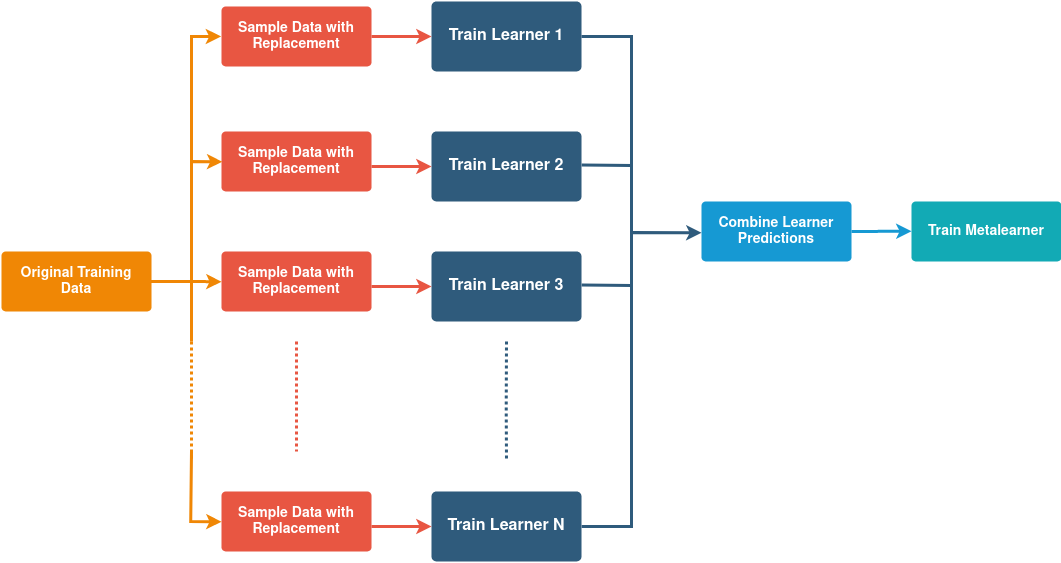}
    \end{adjustbox}
    \caption{Here we describe the procedure used to train the bagging ensemble. The original training dataset is sampled \textit{N} times, with replacement, to create a unique training dataset for each of the \textit{N} learners in the ensemble. These learners are then all trained, a process which can be parallelized due to the mutual independence of the learners. All of the trained learners are then used to classify the training data and their outputs are combined to create the training data for the metalearner, which is then trained.}
    \label{fig:1}
\end{figure}

A bagging ensemble utilizes multiple high accuracy learners and combines their classification outputs using weighted voting to build a powerful ensemble with higher accuracy than any of the constituent classifiers. The training methodology of the ensemble is shown in Figure \ref{fig:1}, where each learner is assigned its own dataset, sampled with replacement from the original training dataset, thereby introducing diversity into the ensemble. Each dataset may contain duplicate data points due to the replacement in the sampling process and there may also be overlap between the different datasets. This process helps prevent the ensemble from overfitting the original dataset and is called bootstrap-aggregating (aka bagging). It also allows the learners to be trained in parallel as they are mutually independent.

\begin{figure}[t!]
\begin{adjustbox}{center}
    \centering
    \includegraphics[width=\columnwidth]{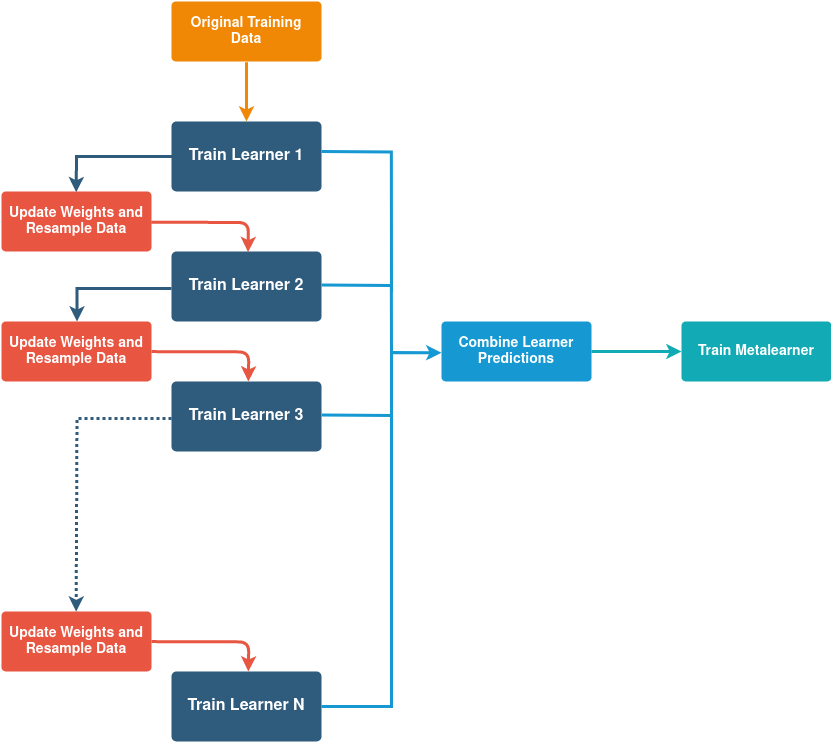}
    \end{adjustbox}
    \caption{Here we describe the procedure used to train the boosting ensemble. The initial learner in the ensemble is trained on the original training data and is used to perform the classification. The weights of the datapoints that are correctly and incorrectly classified are decreased and increased, respectively. This allows the misclassified datapoints to appear more frequently in the subsequent samples of the training data. After resampling with the new weights, the next learner is trained on this new dataset. This process is repeated until either the limit on ensemble size is reached, or until a learner is either entirely correct or incorrect. Next, all of the learners in the ensemble are used to classify the training data and their outputs are combined to create the training data for the metalearner, which is then trained.}
    \label{fig:2}
\end{figure}

A boosting ensemble utilizes multiple weak individual learners (with validation accuracies slightly better than 50\%) and combines them using weighted voting to build a powerful ensemble. The strength of the ensemble comes from the mechanism through which the datasets each learner is trained on are generated, and is illustrated in Figure \ref{fig:2}. The initial learner is trained on the specified dataset and is used to classify the points in the dataset. Each data point is assigned an equal weight at the start of the training and thereafter the weights of the correctly and incorrectly classified data points are decreased and increased, respectively. The dataset is then resampled using the new weights, thereby emphasizing the datapoints that are difficult to classify correctly. This new dataset is given to the next learner and this process is repeated until a classifier which is entirely correct or entirely incorrect is obtained. 

One of the most influential aspects of ensemble methods is the manner in which the predictions of all of the individual models in the ensemble are combined to produce a single output. One of the most common methods is a weighted voting scheme whereby the outputs of each of the ensemble's constituent models are considered and the degree to which they contribute is decided by the weight assigned to the model \citep{dietterich}. As this choice of weights for each of the models in the ensemble is extremely important, we utilized a metalearner that learns the optimal weights to maximize the performance of the ensemble. We create the labelled training data for the metalearner using both the predictions of the individual models on the original training data and the true class labels of the data points. Once trained, the metalearner is capable of combining the predictions of all of the models to produce the most accurate ensemble possible from the data. In this work we select a logistic regression model to use as our metalearner. 

\begin{figure}[!t]

    \begin{adjustbox}{center}
        \subfloat[A trainable block in the QAUM circuit contains sequential $R_z$, $R_x$, and $R_y$ gates.]{
            \includegraphics[width=\columnwidth]{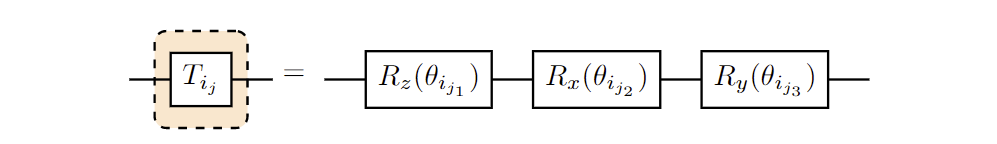}
        }
    \end{adjustbox}

    \begin{adjustbox}{width=\columnwidth, center}
        \subfloat[A layer in the QAUM circuit consists of alternating $R_z$ encoding gates and trainable blocks.]{
            \includegraphics[width=\columnwidth]{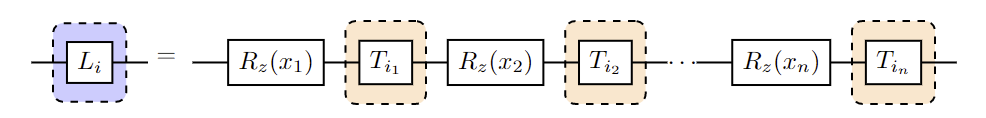}
        }
    \end{adjustbox}

    \begin{adjustbox}{center}
        \subfloat[A full QAUM circuit of $\mathrm{depth}=d$.]{
            \includegraphics[width=\columnwidth]{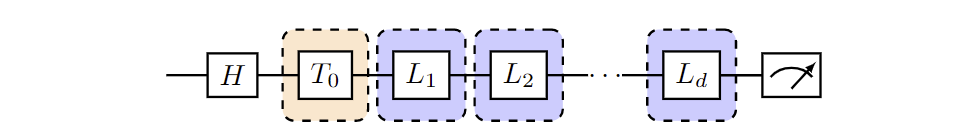}    
        }
    \end{adjustbox}

    \caption{The QAUM circuit from \citep{mo} utilizes an alternating structure of single-qubit data encoding gates and trainable layers of parameterized single-qubit gates. This structure facilitates an increase in model expressivity by increasing the number of layers in the circuit. The structure of the trainable blocks, circuit layers, and complete QAUM circuit can be seen in (a), (b), and (c), respectively.} 
    \label{fig:3}
\end{figure}

\section{Methodology}\label{section:3}

\subsection{Quantum Classifier}

The selection of the individual classifiers that make up the ensemble has a strong effect on the overall success of the ensemble techniques \citep{dietterich}. In this work, the ensembles consist of QAUM circuits, whose design was proposed in \cite{mo}. QAUM circuits, seen in Figure \ref{fig:3}, utilize only a single qubit and are composed of layers of parameterized \textit{Z}, \textit{X}, and \textit{Y}-rotation gates. These rotations make up the trainable blocks, and are interlaced with \textit{Z}-rotation gates to encode the data features \citep{mo}. This choice of gates allows efficient access to both the group and Fourier spaces, and the first degree multi-dimensional Fourier series of the dataset can be expressed by this architecture \citep{mo}.

\begin{figure}[!t]

    \begin{adjustbox}{center}
        \subfloat[The original gate selection for the trainable layers devised in \cite{mo}.]{
            \includegraphics[width=\columnwidth]{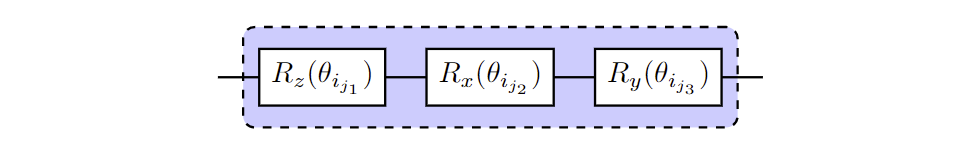}
        }
    \end{adjustbox}

    \begin{adjustbox}{center}
        \subfloat[The modified gate selection used for the trainable layers in this work.]{
            \includegraphics[width=\columnwidth]{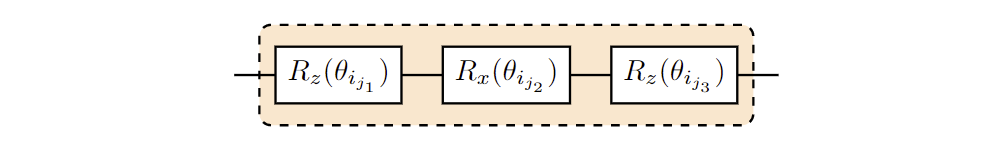}
        }
    \end{adjustbox}

    \caption{In this work, we modify the gate selection in the trainable layers of the QAUM circuit from \cite{mo}. By utilizing the \textit{ZXZ}-rotations in (b), rather than the \textit{ZXY}-rotations in (a), the optimizer is able to more easily converge on the optimal solution thanks to a simpler gate composition. The trainable layer shown in (b) is still capable of efficiently accessing the Fourier space as it can turn any initial state into any final state \citep{nielsen}.}
    \label{fig:4}
\end{figure}

In \cite{Schuld} it was shown that quantum models can naturally be expressed as a partial Fourier series in the data and that the accessible spectrum of frequencies is dictated by the nature of the data encoding gates utilized in the circuit. Once a quantum model has access to a sufficiently large spectrum of frequencies, it becomes a universal function approximator for periodic functions \citep{Schuld}. 

The choice of rotation gates in the trainable layers used in this work was modified so that the optimizer can more easily train the parameters for these gates. The original structure of \textit{Z}, \textit{X}, and \textit{Y}-rotations in Figure \ref{fig:4}(a) was changed to the \textit{Z}, \textit{X}, and \textit{Z}-rotations shown in Figure \ref{fig:4}(b). As the \textit{Y}-rotation in the original circuit is composed of both an \textit{X} and \textit{Z}-rotation, changing it to a \textit{Z}-rotation allows the optimizer to more easily train the parameters of this gate to minimise the loss function. Additionally, this modification has no effect on the accessibility of the group and Fourier spaces as the rotation \textit{ZXZ} can still convert any initial state $\ket{\psi}$ into any final state $\ket{\phi}$ \citep{nielsen}.

\subsection{Datasets}

The main dataset used in this work is the HTRU 2 Pulsar dataset \citep{htru}. This is the binary classification of 8-dimensional input as either radio frequency interference (RFI) or a pulsar star, an important problem for the analysis of gravitational waves and other applications \citep{mo}.

To test whether the performance trends for the classifiers are consistent across different problems, we also utilized a subset of MNIST Digits \citep{mnist} containing the digits 8 and 9, and the Wisconsin Breast Cancer dataset \citep{uci}. In all cases we tested using only 8-dimensional input, so principal component analysis was used to perform dimensionality reduction where needed.

For the purposes of result validation we conducted a paired t-test to compare the classification models. We performed 5$\times$2 cross-validation on the datasets, whereby 5 different samples of the dataset are taken, shuffled, and then split for use in training and validation for each of the models.

\subsection{Optimization and Implementation}

The training of the quantum models which make up the ensembles was performed using the constrained optimization by linear approximation (COBYLA) optimizer \citep{powell_1998}, a gradient-free optimizer which serves our purposes well as the quantum circuits are executed on a noise-free classical simulator.

The implementation of all of the ensembles, as well as the original QAUM classifier from \cite{mo}, can be found at \cite{mcfarthing}. This was accomplished using the Julia programming language \citep{Julia-2017} and the Yao \citep{Yao} library was used for the creation and execution of the quantum variational circuits making up the ensembles' constituent models. 

\subsection{Testing}

Due to the differences between the bagging and boosting ensembles implemented, specific hyper-parameters were selected for each. The bagging algorithm works best with accurate base learners so we trained the learners for a large number of iterations and used only a few learners in the ensemble to keep training overhead down. We tested a bagging ensemble of $7$ QAUM classifiers, each with $2$ embedding layers, and each learner was trained for $1714$ iterations. This ensemble was then compared to a single QAUM classifier, also with $2$ embedding layers, trained for $12 000$ iterations. 

On the other hand, the boosting algorithm relies on learners which do not have high accuracies individually so the number of embedding layers in the quantum circuit was set to $1$ and each learner was only trained for $1$ iteration. This ensemble was then compared to a single QAUM classifier, with $2$ embedding layers, trained for $100$ iterations.

\section{Results}\label{section:4}

\begin{figure}[t!]
\begin{adjustbox}{center}
    \centering
    \includegraphics[width=\columnwidth]{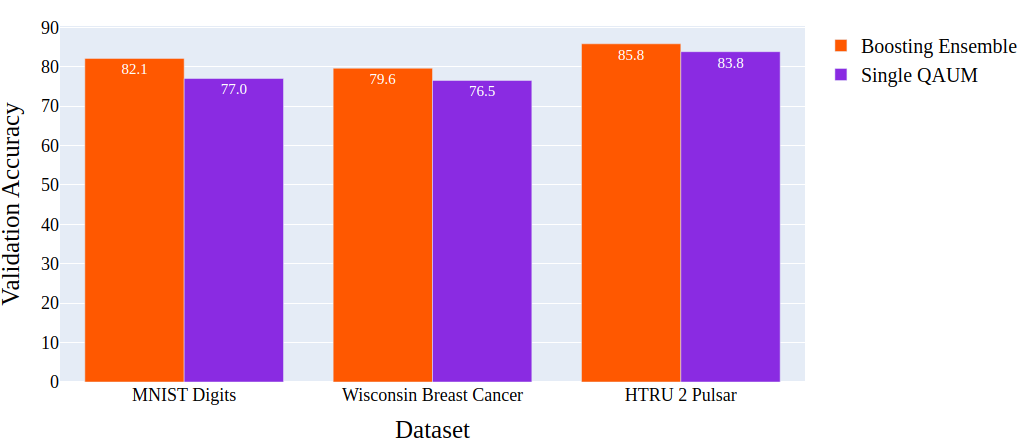}
    \end{adjustbox}
    \caption{The mean validation accuracies of the boosting ensemble (orange) and original QAUM classifier (purple) are shown. Testing was performed using 5$\times$2 cross-validation for each dataset. We find that the boosting ensemble outperforms the original QAUM classifier for the HTRU 2 Pulsar dataset \citep{htru}, as well as the MNIST Digits and Wisconsin Breast Cancer datasets \citep{mnist, uci}.}
    \label{fig:5}
\end{figure}

\begin{table}[t!]
\normalsize
    \begin{adjustbox}{center}
        \resizebox{\columnwidth}{!}{
            \begin{tabular}[h]{|p{1.5cm}||p{1.7cm}|p{1.2cm}|p{1.6cm}||p{1.5cm}|p{1.5cm}|p{1.5cm}|p{1.4cm}| }
                 \hline
                \multicolumn{8}{|c|}{Validation Accuracy and Training Time for each Dataset} \\
                \hline
                Type of Classifier& Number of Learners & Circuit Depth & Epochs per Learner & Digits Dataset Accuracy & Cancer Dataset Accuracy& Pulsar Dataset Accuracy& Average time to Train\\
                \hline
                Boosting Ensemble&$27$&$1$&$1$&$82.1\pm5.4$&$79.6\pm9.2$&$85.8\pm4.8$&$1.7$ seconds\\
                \hline
                Single QAUM&$1$&$2$&$100$&$77.0\pm5.6$&$76.5\pm9.0$&$83.8\pm5.7$&$3.6$ seconds\\
                \hline
            \end{tabular}
        }
    \end{adjustbox}
    \caption{The classifier details and validation accuracies for the testing performed on the HTRU 2 Pulsar, MNIST Digits, and Wisconsin Breast Cancer datasets used \citep{htru, mnist, uci}. The number of learners for the boosting ensemble is taken as the average over all 3 datasets, as the ensemble determines this number based on the difficulty of the dataset. The boosting ensemble achieves higher classification accuracies than the single QAUM classifier in each of the classification problems and is more than twice as fast to train.}
    \label{table:1}
\end{table}

To verify the significance of the results obtained from our testing, a paired t-test with a threshold of $0.05$ was utilized in conjunction with 5$\times$2 cross-validation. In all tests the p-value was significantly lower than the threshold, thereby confirming the legitimacy of the results.

Figure \ref{fig:5} and Table \ref{table:1} show that the boosting ensemble outperformed the single QAUM classifier for each of the datasets used. The boosting ensemble had validation accuracies that were \textbf{$2$\%}, \textbf{$5.1$\%}, and \textbf{$3.1$\%} higher than those of the single QAUM classifier for the Pulsar, Digits, and Cancer datasets, respectively. Additionally, the time taken to train the boosting ensemble was less than half that of the single QAUM classifier. The validation accuracies for each of the samples in the 5$\times$2 cross-validation can be found in Appendix \ref{secA1}. 

Figure \ref{fig:6} and Table \ref{table:2} show that the bagging ensemble outperformed the single QAUM classifier for each of the datasets used. The bagging ensemble had validation accuracies that were \textbf{$1.6$\%}, \textbf{$3.1$\%}, and \textbf{$0.6$\%} higher than those of the single QAUM classifier for the Pulsar, Digits, and Cancer datasets, respectively. The validation accuracies for each of the samples in the 5$\times$2 cross-validation can be found in Appendix \ref{secA2}.

It is noticeable that the accuracies achieved by the boosting ensemble in Table \ref{table:1} are lower than those of the bagging ensemble in Table \ref{table:2}. Additionally, the duration of training was far lower for the boosting ensemble than for the bagging ensemble, thus raising the question of the use case for each of these ensembles.

\begin{figure}[t!]
\begin{adjustbox}{center}
    \includegraphics[width=\columnwidth]{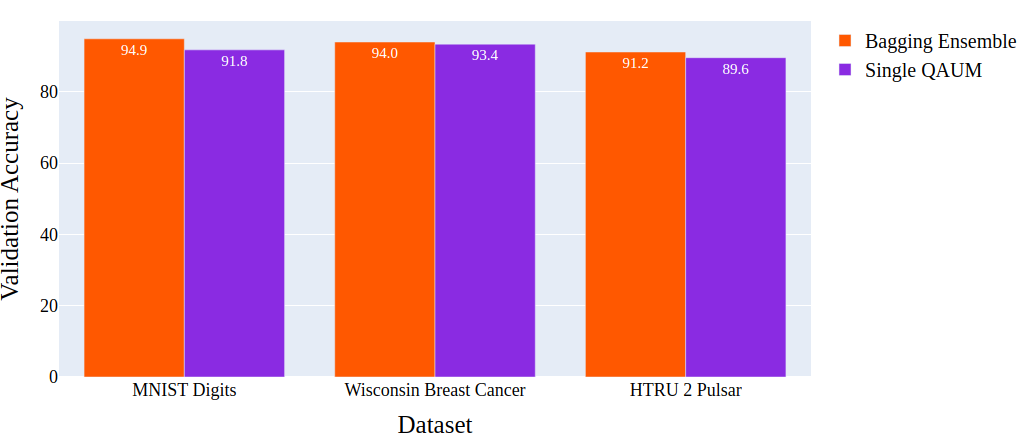}
    \end{adjustbox}
    \caption{The mean validation accuracies of the bagging ensemble (orange) and original QAUM classifier (purple) are shown. Testing was performed using 5$\times$2 cross-validation for each dataset. We find that the bagging ensemble outperforms the original QAUM classifier for the HTRU 2 Pulsar dataset \citep{htru}, as well as the MNIST Digits and Wisconsin Breast Cancer datasets \citep{mnist, uci}.}
    \label{fig:6}
\end{figure}

\begin{table}[t]
\normalsize
    \begin{adjustbox}{center}
        \resizebox{\columnwidth}{!}{
            \begin{tabular}[h]{|p{1.5cm}||p{1.7cm}|p{1.2cm}|p{1.6cm}|| p{1.5cm}|p{1.5cm}|p{1.5cm}|p{1.4cm}| }
                 \hline
                 \multicolumn{8}{|c|}{Validation Accuracy and Training Time for each Dataset} \\
                 \hline
                 Type of Classifier& Number of Learners & Circuit Depth & Epochs per Learner & Digits Dataset Accuracy & Cancer Dataset Accuracy& Pulsar Dataset Accuracy& Average time to Train\\
                 \hline
                 Bagging Ensemble&$7$&$2$&$1714$&$94.9\pm2.4$&$94.0\pm2.4$&$91.2\pm3.8$&$207.6$ seconds\\
                 \hline
                 Single QAUM&$1$&$2$&$12000$&$91.8\pm2.1$&$93.4\pm2.9$&$89.6\pm5.8$&$349.3$ seconds\\
                 \hline
            \end{tabular}
        }
    \end{adjustbox}
    \caption{The classifier details and validation accuracies for the testing performed on the HTRU 2 Pulsar, MNIST Digits, and Wisconsin Breast Cancer datasets used \citep{htru, mnist, uci}. The bagging ensemble achieves higher classification accuracies than the single QAUM classifier in each of the classification problems. Similarly to the boosting ensemble, the bagging ensemble is also quicker to train than the comparative QAUM classifier.}
    \label{table:2}
\end{table}

In Figure \ref{fig:7}, the validation accuracies for both types of ensembles are shown in relation to training length per learner and it is clear that the boosting ensemble is able to achieve a decent level of accuracy with almost no training at all. However, as the number of training epochs is increased, the bagging ensemble overtakes the boosting ensemble for higher overall accuracy.

\section{Conclusions}
\label{section:conclusion}

In this work we explored the use of classical ensemble techniques, specifically bagging and boosting, as a means to expand the classification power of the QAUM classifier from \cite{mo}.

\begin{figure}[t!]
\begin{adjustbox}{center}
    \includegraphics[width=\columnwidth]{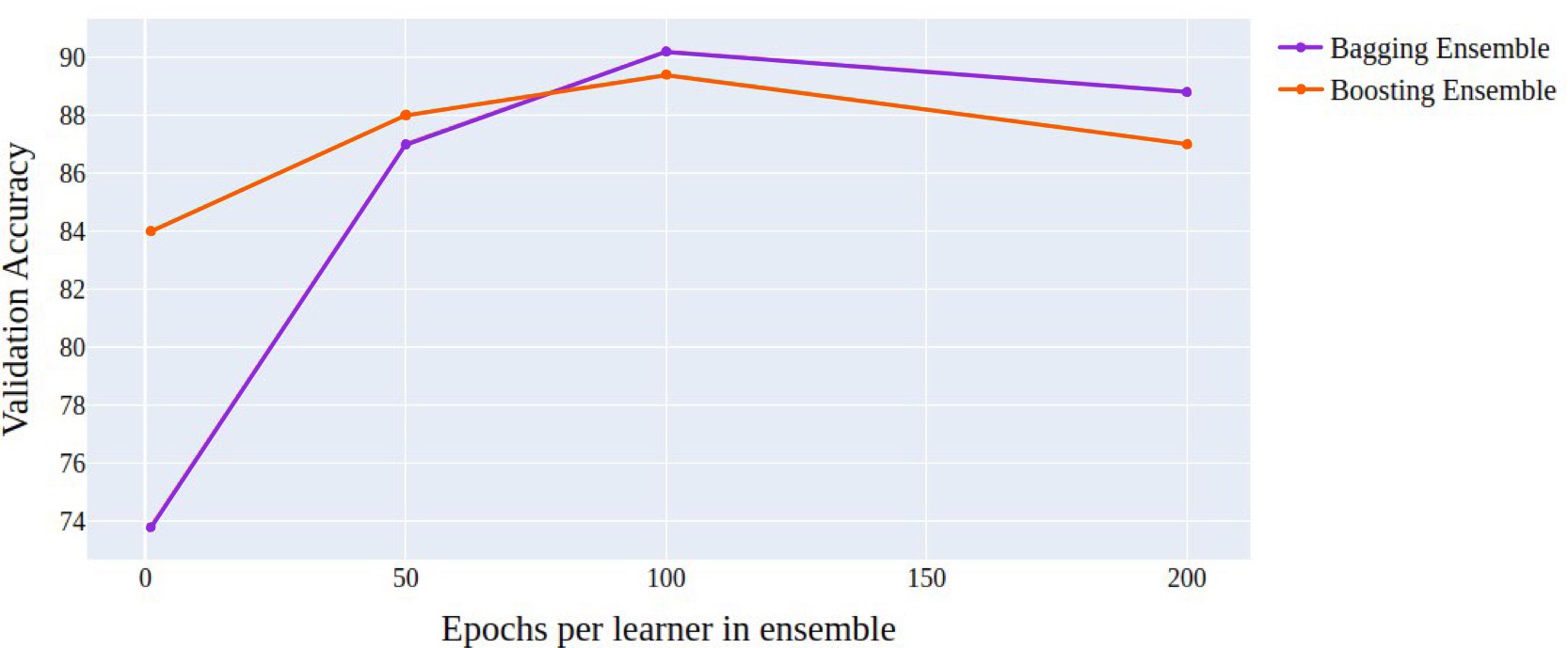}
    \end{adjustbox}
    \caption{Validation accuracy of the bagging and boosting ensembles on the HTRU 2 Pulsar \citep{htru} dataset with 100 samples. The trade-off between training time requirements and overall accuracy for the two types of ensembles can be seen.}
    \label{fig:7}
\end{figure}

Through testing on multiple datasets \citep{htru, mnist, uci}, we have shown that both boosting and bagging ensembles outperform the original QAUM circuit. This is shown in Figures \ref{fig:5} and \ref{fig:6}. Additionally, we analyzed the differences between the results of the boosting and bagging ensemble, shown in Figure \ref{fig:7}, and propose that the boosting ensemble of QAUM classifiers is applicable to problems where training time is prioritised over accuracy. This is due to its ability to achieve a reasonable level of performance very quickly. Furthermore, we propose that the bagging ensemble be used when accuracy is of the utmost performance as it is capable of generalizing better than the boosting ensemble with sufficient training.

The application of these classical ensemble techniques to the QAUM single-qubit circuits was investigated as an alternative to other multi-qubit approaches and the results obtained in this work show that this is promising as improvements in classification accuracy can be achieved with relatively shallow circuits. The original QAUM circuit \citep{mo} was designed to achieve higher accuracy with increasing circuit depth. However, due to the low coherence time of current devices, this would not be practical. Through the use of the techniques in this work, a powerful ensemble can be built from shallow single-qubit circuits which can be run on existing quantum hardware. 

The testing of this approach on real quantum hardware can be done in future research, as well as investigations regarding the noise resilience of these shallow ensembles compared to the original QAUM classifier.

\bibliography{sn-bibliography}

\begin{thebibliography}{18}
\providecommand{\natexlab}[1]{#1}
\providecommand{\url}[1]{{#1}}
\providecommand{\urlprefix}{URL }
\providecommand{\doi}[1]{\url{https://doi.org/#1}}
\providecommand{\eprint}[2][]{\url{#2}}
 \bibcommenthead

\bibitem[{Bezanson et~al(2017)Bezanson, Edelman, Karpinski, and
  Shah}]{Julia-2017}
Bezanson J, Edelman A, Karpinski S, et~al (2017) Julia: A fresh approach to
  numerical computing. SIAM review 59(1):65--98. \doi{10.1137/141000671}

\bibitem[{Dietterich(2000)}]{dietterich}
Dietterich TG (2000) Ensemble methods in machine learning. In: International
  workshop on multiple classifier systems, Springer, pp 1--15,
  \doi{10.1007/3-540-45014-9_1}

\bibitem[{Dua and Graff(2017)}]{uci}
Dua D, Graff C (2017) {UCI} machine learning repository.
  \urlprefix\url{http://archive.ics.uci.edu/ml}

\bibitem[{Kordzanganeh et~al(2021)Kordzanganeh, Utting, and Scaife}]{mo}
Kordzanganeh M, Utting A, Scaife A (2021) Quantum machine learning for radio
  astronomy. arXiv preprint arXiv:211202655 \doi{10.48550/arXiv.2112.02655}

\bibitem[{LeCun and Cortes(2010)}]{mnist}
LeCun Y, Cortes C (2010) {MNIST} handwritten digit database.
  http://yann.lecun.com/exdb/mnist/,
  \urlprefix\url{http://yann.lecun.com/exdb/mnist/}

\bibitem[{Li et~al(2023)Li, Huang, Hou, Li, Wang, and Bayat}]{li2023ensemble}
Li Q, Huang Y, Hou X, et~al (2023) Ensemble-learning variational
  shallow-circuit quantum classifiers. arXiv preprint arXiv:230112707
  \doi{10.48550/arXiv.2301.12707}

\bibitem[{Luo et~al(2020)Luo, Liu, Zhang, and Wang}]{Yao}
Luo XZ, Liu JG, Zhang P, et~al (2020) Yao. jl: Extensible, efficient framework
  for quantum algorithm design. Quantum 4:341. \doi{10.22331/q-2020-10-11-341}

\bibitem[{Lyon et~al(2016)Lyon, Stappers, Cooper, Brooke, and Knowles}]{htru}
Lyon RJ, Stappers B, Cooper S, et~al (2016) Fifty years of pulsar candidate
  selection: from simple filters to a new principled real-time classification
  approach. Monthly Notices of the Royal Astronomical Society
  459(1):1104--1123. \doi{10.1093/mnras/stw656}

\bibitem[{Macaluso et~al(2020)Macaluso, Clissa, Lodi, and Sartori}]{macaluso}
Macaluso A, Clissa L, Lodi S, et~al (2020) Quantum ensemble for classification.
  arXiv preprint arXiv:200701028 \doi{10.48550/arXiv.2007.01028}

\bibitem[{McClean et~al(2018)McClean, Boixo, Smelyanskiy, Babbush, and
  Neven}]{mcclean}
McClean JR, Boixo S, Smelyanskiy VN, et~al (2018) Barren plateaus in quantum
  neural network training landscapes. Nature communications 9(1):1--6.
  \doi{10.1038/s41467-018-07090-4}

\bibitem[{McFarthing(2022)}]{mcfarthing}
McFarthing S (2022) Classical ensembles of single-qubit quantum variational
  circuits for classification.
  \urlprefix\url{https://github.com/shanemcfarthing/Classical-Ensembles-of-Single-Qubit-Quantum-Classifiers.git}

\bibitem[{Nielsen and Chuang(2002)}]{nielsen}
Nielsen MA, Chuang I (2002) Quantum computation and quantum information.
  American Association of Physics Teachers, \doi{10.1119/1.1463744}

\bibitem[{Powell(1998)}]{powell_1998}
Powell MJ (1998) Direct search algorithms for optimization calculations. Acta
  numerica 7:287--336. \doi{10.1017/S0962492900002841}

\bibitem[{Qin et~al(2022)Qin, Liang, Cheng, Kogge, and Shi}]{qin}
Qin R, Liang Z, Cheng J, et~al (2022) Improving quantum classifier performance
  in nisq computers by voting strategy from ensemble learning. arXiv preprint
  arXiv:221001656 \doi{10.48550/arXiv.2210.01656}

\bibitem[{Schuld and Petruccione(2018{\natexlab{a}})}]{Schuld2018}
Schuld M, Petruccione F (2018{\natexlab{a}}) Quantum ensembles of quantum
  classifiers. Scientific reports 8(1):1--12. \doi{10.1038/s41598-018-20403-3}

\bibitem[{Schuld and Petruccione(2018{\natexlab{b}})}]{petruccione}
Schuld M, Petruccione F (2018{\natexlab{b}}) Supervised learning with quantum
  computers, vol~17. Springer, \doi{10.1007/978-3-319-96424-9}

\bibitem[{Schuld et~al(2015)Schuld, Sinayskiy, and Petruccione}]{schuld2015}
Schuld M, Sinayskiy I, Petruccione F (2015) An introduction to quantum machine
  learning. Contemporary Physics 56(2):172--185.
  \doi{10.1080/00107514.2014.964942}

\bibitem[{Schuld et~al(2021)Schuld, Sweke, and Meyer}]{Schuld}
Schuld M, Sweke R, Meyer JJ (2021) Effect of data encoding on the expressive
  power of variational quantum-machine-learning models. Physical Review A
  103(3):032,430. \doi{10.1103/PhysRevA.103.032430}

\end{thebibliography}


\appendix

\begin{figure}[b]
\begin{adjustbox}{center}
    \includegraphics[width=\columnwidth]{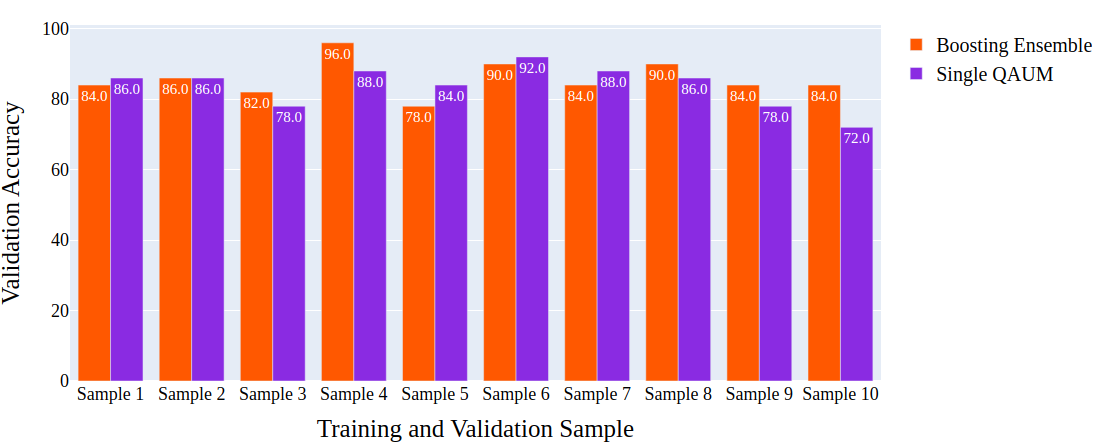}
    \end{adjustbox}
    \caption{Validation accuracies of the boosting ensemble and single QAUM classifier on the HTRU 2 Pulsar \citep{htru} dataset for each of the samples used in the 5$\times$2 cross-validation. In each group of bars the accuracies of the boosting ensemble and single QAUM classifier are represented by the left and right bars, respectively.}
    \label{fig:8}
\end{figure}

\section{Cross-validation results for Boosting Ensemble vs. Single QAUM Classifier}\label{secA1}

\begin{figure}[ht]
\begin{adjustbox}{center}
    \includegraphics[width=\columnwidth]{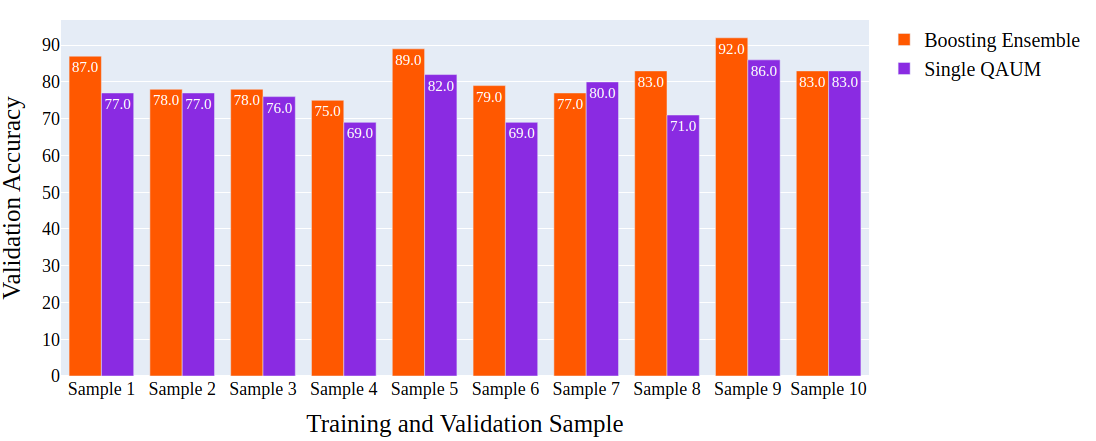}
    \end{adjustbox}
    \caption{Validation accuracies of the boosting ensemble and single QAUM classifier on the MNIST Digits \citep{mnist} dataset for each of the samples used in the 5$\times$2 cross-validation. In each group of bars the accuracies of the boosting ensemble and single QAUM classifier are represented by the left and right bars, respectively.}
    \label{fig:9}
\end{figure}

\begin{figure}[ht]
\begin{adjustbox}{center}
    \includegraphics[width=\columnwidth]{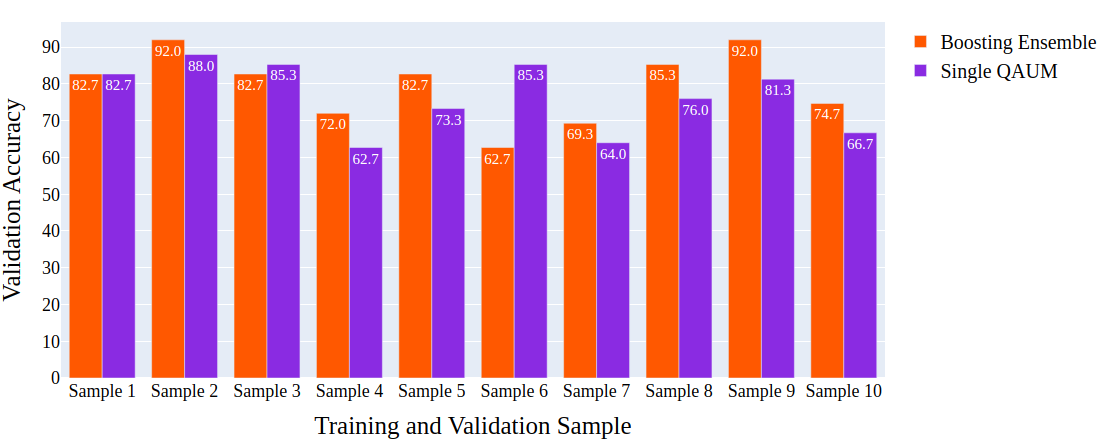}
    \end{adjustbox}
    \caption{Validation accuracies of the boosting ensemble and single QAUM classifier on the Wisconsin Breast Cancer \citep{uci} dataset for each of the samples used in the 5$\times$2 cross-validation. In each group of bars the accuracies of the boosting ensemble and single QAUM classifier are represented by the left and right bars, respectively.}
    \label{fig:10}
\end{figure}

The validation accuracies for each of the dataset samples used in the cross-validation are shown in Figures \ref{fig:8}, \ref{fig:9}, and \ref{fig:10}. As mentioned previously, the 3 datasets used in testing were HTRU 2 Pulsar, MNIST Digits, and Wisconsin Breast Cancer \citep{htru, mnist, uci}.

\FloatBarrier

\section{Cross-validation graphs for Bagging Ensemble vs. Single QAUM Classifier}\label{secA2}

\begin{figure}[ht]
\begin{adjustbox}{center}
    \includegraphics[width=\columnwidth]{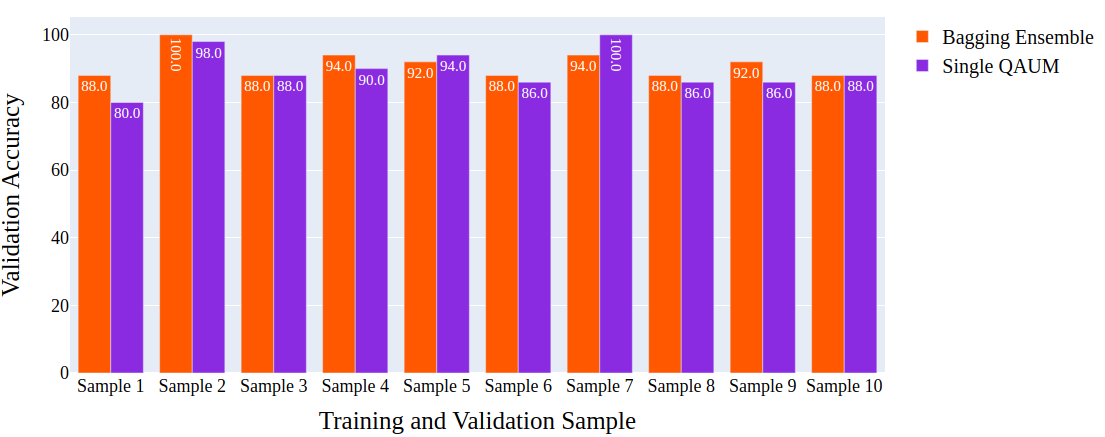}
    \end{adjustbox}
    \caption{Validation accuracies of the bagging ensemble and single QAUM classifier on the HTRU 2 Pulsar \citep{htru} dataset for each of the samples used in the 5$\times$2 cross-validation. In each group of bars the accuracies of the bagging ensemble and single QAUM classifier are represented by the left and right bars, respectively.}
    \label{fig:11}
\end{figure}

\begin{figure}[b]
\begin{adjustbox}{center}
    \includegraphics[width=\columnwidth]{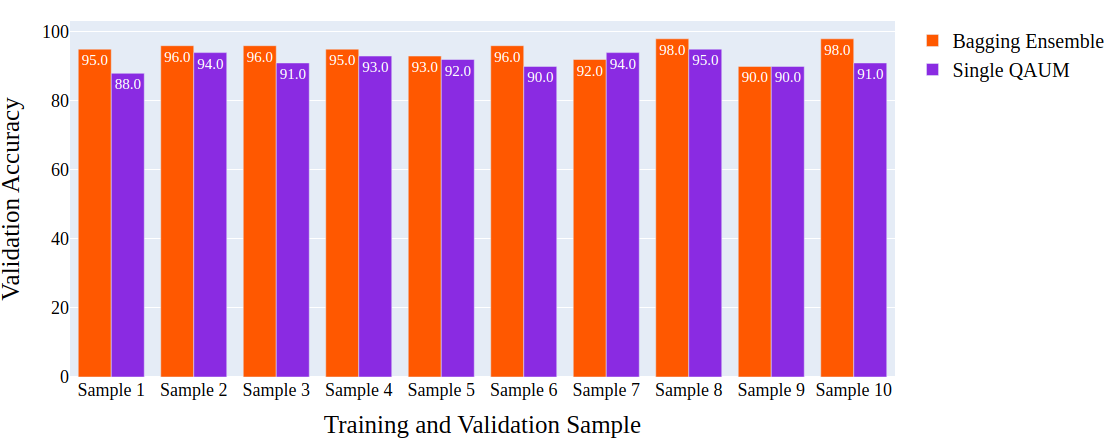}
    \end{adjustbox}
    \caption{Validation accuracies of the bagging ensemble and single QAUM classifier on the MNIST Digits \citep{mnist} dataset for each of the samples used in the 5x2 cross-validation. In each group of bars the accuracies of the bagging ensemble and single QAUM classifier are represented by the left and right bars, respectively.}
    \label{fig:12}
\end{figure}

\begin{figure}[ht]
\begin{adjustbox}{center}
    \includegraphics[width=\columnwidth]{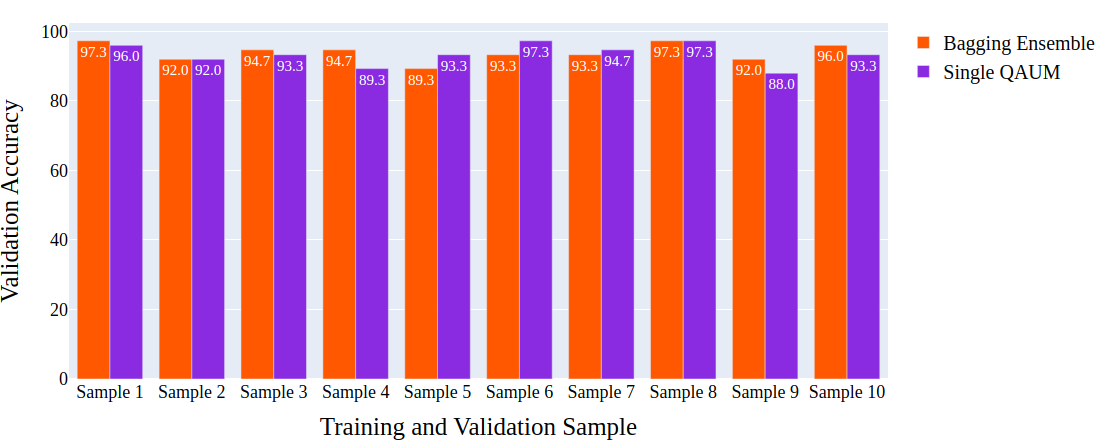}
    \end{adjustbox}
    \caption{Validation accuracies of the bagging ensemble and single QAUM classifier on the Wisconsin Breast Cancer \citep{uci} dataset for each of the samples used in the 5$\times$2 cross-validation. In each group of bars the accuracies of the bagging ensemble and single QAUM classifier are represented by the left and right bars, respectively.}
    \label{fig:13}
\end{figure}

The validation accuracies for each of the dataset samples used in the cross-validation are shown in Figures \ref{fig:11}, \ref{fig:12}, and \ref{fig:13}. As mentioned previously, the 3 datasets used in testing were HTRU 2 Pulsar, MNIST Digits, and Wisconsin Breast Cancer \citep{htru, mnist, uci}. 

\FloatBarrier

\FloatBarrier




\end{document}